\documentclass[fleqn,12pt,twoside]{article}
\usepackage{espcrc1}

\usepackage{graphicx}

\usepackage[figuresright]{rotating}

\newcommand{\AmS}{{\protect\the\textfont2
  A\kern-.1667em\lower.5ex\hbox{M}\kern-.125emS}}

\title{Structure and texture of the quark mass matrix}

\author{Maritza de Coss
\address{Departamento de F{\'\i}sica Aplicada,\\
        Centro de Investigaci\'on y de Estudios Avanzados del
        Instituto Polit\'ecnico Nacional,\\
        Unidad M\'erida, A. P. 73 Cordemex 97310, M\'erida, Yucat\'an,
        M\'exico.}
and Rodrigo Huerta\addressmark }

\begin{document}

\maketitle

\begin{abstract}

Starting from a weak basis in which the up (or down) quark matrix is
diagonal, 
we obtain an exact set of equations for the quark mass matrix elements
in terms of known observables.
We make a numerical analysis of the down (up) quark mass
matrix. Using the data available for the quark masses and mixing angles
at different energy scales,
we found a numerical expression for these matrices.
We suggest that it is not possible to have
an specific texture from this analysis. We also examine the
most general case when the complex phases are introduced in the mass matrix.
We find the numerical value for these phases as a function of $\delta$,
the CP-violationg phase.

\bigskip
PACS number(s): 12.15.Ff

\end{abstract}

\section{Introduction}

The known observables in the Yukawa sector of the Standard Model
(SM) are six quark mass eigenvalues plus the four parameters
of the  Cabibbo-Kobayashi-Maskawa (CKM) mixing matrix.
These parameters are related to the quark mixing matrices
through the diagonalization procedure.
The two quark mass matrices with three generations depend on 36
parameters in general (18 each).
There are many possibilities in the literature that propose
different structures for these matrices.
In particular, there are studies considering the mass matrix as
hermitian, symmetric~\cite{fritzsch78,georgi79,ramond93}
or non-hermitian~\cite{branco94}.
To reduce the number of parameters, zeros are introduced in the
mass matrices leading to the terminology of texture~\cite{ramond93}.
Depending on the particular texture it is possible to find relations
among masses and mixing angles.
It would be of interest to explore what is the texture at which the
present data is pointing.

In this work we obtain a set of exact equations relating the
mass matrix parameters and the known observables.
We perform a numerical analysis with those algebraic equations,
using the available data, and we find the structure of the quark
mass matrices. We do this analysis for different energy scales
using the renormalization group equation to find the evolution
of the known observables.
In the next section we obtain the exact set of equations from which
we observe the relation among the measured quantities and the
quark mass matrix elements, including the phases of the latter.
In section III, we find the evolution of these matrix elements.
From this analysis we find their hierarchy. Finally, in section IV
we sumarize our results.

\section{Expressions for the mass matrix elements}

We start from the terms of the mass and the weak charged-current of
the standard model Lagrangian which are important to us,

\begin{equation}
{\cal L}_{mass}
=\bar{u}_L^{\prime}{\mathbf m}_uu_R^{\prime}
+\bar{d}_L^{\prime}{\mathbf m}_dd_R^{\prime}
+g_L\bar{u}_L^{\prime}{\mathbf W}^{\dagger}_Ld_L^{\prime}+c.h.,
\end{equation}

\noindent
where the quark mass matrices are general complex matrices whose
size depends on the number of generations. We can rotate the weak
states to get a diagonal mass matrix by making the usual
transformations~\cite{wein},

\begin{equation}
\begin{array}{lll}
\vspace{2mm}
{\cal L}_{mass}
&=&\bar{u}_L^{\prime}{\mathbf R}_L^u({\mathbf R}_L^u)^{\dagger}{\mathbf m}_u
({\mathbf S}_R^u)^{\dagger}{\mathbf S}_R^uu_R^{\prime}
+\bar{d}_L^{\prime}{\mathbf R}_L^d({\mathbf R}_L^d)^{\dagger}{\mathbf m}_d
({\mathbf S}_R^d)^{\dagger}{\mathbf S}_R^dd_R^{\prime}\\
\vspace{2mm}
&&+g_L\bar{u}_L({\mathbf R}^u_L)^{\dagger}{\mathbf W}_L^{\dagger}{\mathbf R}^d_Ld_L
+c.h.\\
&=&\bar{u}_L{\mathbf M}_uu_R+\bar{d}_L{\mathbf M}_dd_R
+g_L\bar{u}_L{\mathbf W}_L^{\dagger}{\mathbf V}_{CKM}d_L+c.h.
\end{array}
\end{equation}

\noindent
where the unitary matrices ${\mathbf R}^{u,d}_L$ and  ${\mathbf S}^{u,d}_R$
rotate the weak basis to the mass eigenstates,

\begin{equation}
\begin{array}{cc}
\vspace{2mm}
u_L^{\prime}={\mathbf R}_L^uu_L,&
d_L^{\prime}={\mathbf R}_L^dd_L\\
u_R^{\prime}={\mathbf S}_R^uu_R,&
d_R^{\prime}={\mathbf S}_R^dd_R.\\
\end{array}
\end{equation}

In this way we obtain the biunitary transformations which transform the mass
matrices ${\mathbf m}_u$ and ${\mathbf m}_d$ to their diagonal form,

\begin{equation}
\begin{array}{c}
\vspace{2mm}
{\mathbf M}_u=({\mathbf R}_L^u)^{\dagger}{\mathbf m}_u{\mathbf S}_R^u,\\
{\mathbf M}_d=({\mathbf R}_L^d)^{\dagger}{\mathbf m}_d{\mathbf S}_R^d.
\end{array}
\end{equation}

The CKM mixing matrix~\cite{ckm} is then given by

\begin{equation}
{\mathbf V}_{CKM}^L=({\mathbf R}^u_L)^{\dagger}{\mathbf R}^d_L.
\end{equation}

In the case of having a charged right-handed current of the type
$g_R\bar{u}_R^{\prime}{\mathbf W}_R^{\dagger}d_R^{\prime}$
in the Lagrangian~\cite{mohapatra}
(neglecting the possible mixing between the gauge bosons),
 we would have

\begin{equation}
{\mathbf V}_{CKM}^R=({\mathbf S}^u_R)^{\dagger}{\mathbf S}^d_R.
\end{equation}

Without any loss of generality we can have one of the mass matrices
diagonal~\cite{ma91,koide92}. Assume first that

\begin{equation}
{\mathbf m}_u=\left(
\begin{array}{ccc}
m_u&0&0\\
0&m_c&0\\
0&0&m_t
\end{array}
\right),\end {equation}

\noindent
then the matrices ${\mathbf R}_L^u$ and ${\mathbf S}_R^u$ are
equal to the unit matrix. From equations (5) and (6) we obtain

\begin{equation}
\begin{array}{c}
\vspace{2mm}
{\mathbf V}_{CKM}^L={\mathbf R}^d_L,\\
{\mathbf V}_{CKM}^R={\mathbf S}^d_R.
\end{array}
\end{equation}

In this case the diagonal mass matrix $d$ is given by,

\begin{equation}
{\mathbf M}_d=({\mathbf V}_{CKM}^L)^{\dagger}{\mathbf m}_d{\mathbf V}_{CKM}^R.
\end{equation}

One can make ${\mathbf V}_{CKM}^R$ equal to ${\mathbf V}_{CKM}^L$~\cite{ma91} and
$g_L$ equal to $g_R$
assuming {\it manifiest} left-right symmetry in the flavor sector, and then one
can see that we need only to consider a single mixing matrix which is
responsible of the diagonalization,

\begin{equation}
{\mathbf M}_d=({\mathbf V}_{CKM})^{\dagger}{\mathbf m}_d{\mathbf V}_{CKM}.
\end{equation}

This is equivalent to assuming that ${\mathbf m}_d$ is hermitian.
Now we can express the mixing angles in terms of the mass eigenvalues
$m_i$ and the mass matrix elements $m_{ij}$. To this end we take an
hermitian mass matrix and choose the standard form for ${\mathbf V}_{CKM}$,

\begin{equation}
{\mathbf V}_{CKM}=\left(
\begin{array}{ccc}
\vspace{2mm}
c_{12}c_{13}&s_{12}c_{13}&s_{13}e^{-i\delta}\\
\vspace{2mm}
-s_{12}c_{23}-c_{12}s_{23}s_{13}e^{i\delta}&c_{12}c_{23}-s_{12}s_{23}s_{13}e^{i\delta}
&s_{23}c_{13}\\
\vspace{2mm}
s_{12}s_{23}-c_{12}c_{23}s_{13}e^{i\delta}&-c_{12}s_{23}-s_{12}c_{23}s_{13}e^{i\delta}
&c_{23}c_{13}\\
\end{array}\right),
\end{equation}

\noindent
so we have

\begin{equation}
{\mathbf V}_{CKM}
\left(\begin{array}{ccc}
m_1&0&0\\
0&m_2&0\\
0&0&m_3\\
\end{array}\right)
=
\left(\begin{array}{lclcl}
m_{11}                 &&m_{12}e^{i\delta_{12}}&&m_{13}e^{i\delta_{13}}\\
m_{12}e^{-i\delta_{12}}&&m_{22}                 &&m_{23}e^{i\delta_{23}}\\
m_{13}e^{-i\delta_{13}}&&m_{23}e^{-i\delta_{23}}&&m_{33}\\
\end{array}\right)
{\mathbf V}_{CKM}.
\end{equation}

Considering the magnitudes of the elements (2,3), (1,3) and (1,2) on
both sides of this equation we get the mixing angles in exact
form~\cite{rasin98},

\begin{equation}
\frac{s_{23}}{c_{23}}=\frac
{(m_3-m_{11})m_{23}+m_{13}m_{12}}
{(m_3-m_{22})(m_3-m_{11})-m_{12}^2}
\end{equation}
\begin{equation}
\frac{s_{13}}{c_{13}}=
\frac{m_{12}s_{23}+m_{13}c_{23}}
{m_3-m_{11}}
\end{equation}
\begin{equation}
\frac{s_{12}}{c_{12}}=\frac{
m_{12}c_{23}-m_{13}s_{23}}
{(m_2-m_{11})c_{13}+(m_{12}s_{23}+m_{13}c_{23})s_{13}}
\end{equation}

\noindent 
where $s_{ij}=\sin{\theta_{ij}}$, $c_{ij}=\cos{\theta_{ij}}$. 
To obtain the quark masses as functions of the matrix elements we use
the fact that ${\mathbf m}_d$ satisfies the following characteristic equation,

\begin{equation}
\begin{array}{rl}
det({\mathbf m}_d-m{\mathbf 1})=
&-m^3+(m_{11}+m_{22}+m_{33})m^2
-(m_{11}m_{22}+m_{11}m_{33}\\
&+m_{22}m_{33}-m_{23}^2-m_{13}^2-m_{12}^2)m+m_{11}(m_{22}m_{33}
-m_{23}^2)\\
&-m_{12}(m_{12}m_{33}-m_{13}m_{23})
+m_{13}(m_{12}m_{23}-m_{13}m_{22})\\
=&0.
\end{array}
\end{equation}

The eigenvalues $m_i$ also satisfy the equation

\begin{equation}
\begin{array}{rl}
(m_1-m)(m_2-m)(m_3-m)=&
-m^3+(m_1+m_2+m_3)m^2\\
&-(m_1m_2+m_1m_3+m_2m_3)m+m_1m_2m_3\\
=&0.
\end{array}
\end{equation}

After equating the coefficients with the same power of $m$ in (16)
and (17) we get,

\begin{equation}
m_1+m_2+m_3= m_{11}+m_{22}+m_{33}
\end{equation}
\begin{equation}
m_1m_2+m_1m_3+m_2m_3=
m_{11}m_{22}+m_{11}m_{33}+m_{22}m_{33}-m_{23}^2-m_{13}^2-m_{12}^2
\end{equation}
\begin{equation}
\begin{array}{lll}
m_1m_2m_3&=&m_{11}(m_{22}m_{33}-m_{23}^2)
-m_{12}(m_{12}m_{33}-m_{13}m_{23})\\
&&+m_{13}(m_{12}m_{23}-m_{13}m_{22}).
\end{array}
\end{equation}

Going back to eq. (12), we compare the phases of elements
(1,2), (1,3) and (2,3) on both sides of the equation to get the
relations,

\begin{equation}
\tan{(\delta_{12})}=\frac{-(m_3-s_{12}^2m_2-m_1)s_{13}s_{23}\sin{\delta}}
{(m_2-m_1)s_{12}+(m_3-s_{12}^2m_2-m_1)s_{13}s_{23}\cos{\delta}}
\end{equation}
\begin{equation}
\tan{(\delta_{13})}=\frac{(m_3-s_{12}^2m_2-m_1)s_{13}\sin{\delta}}
{(m_2-m_1)s_{12}s_{23}-(m_3-s_{12}^2m_2-m_1)s_{13}\cos{\delta}}
\end{equation}
\begin{equation}
\tan{(\delta_{23})}=\frac{(m_2-m_1)(1+s_{23}^2)s_{12}s_{13}\sin{\delta}}
{[m_3-(s_{12}^2-s_{13}^2)m_1-(1-s_{12}^2s_{13}^2)m_2]s_{23}
-(m_2-m_1)(1-s_{23}^2)s_{12}s_{13}\cos{\delta}}.
\end{equation}

We observe that the phases $\delta_{ij}$ are given in terms of known
observables and that they are independent of the $m_{ij}$.
We consider important to remark that the  set of equations (13-15),
(18-20) and (21-23) are exact relations among quark mass
matrix elements and known observables.

\section{Structure and evolution of the mass matrix elements}

The structure and evolution of the mass matrix can be found by
numerically solving the set of equations obtained in the previous section.
To this end we give in the following the quark mass eigenvalues, the
mixing angles and the $\delta$-phase at different scales.
For the quark mass eigenvalues we use the quantities shown in Table 
1~\cite{koide98}.

\begin{table}[ht!]
\caption{Running quark masses (in units of $GeV$).}
\label{table:1}
\begin{tabular}{ccccccc}
\hline\hline
\emph{Scale}&&\emph{$u$}&&\emph{d}&&\emph{s}\\
\hline
\emph{1 GeV}
&&\emph{$0.00488 \pm 0.00057$}
&&\emph{$0.00981 \pm 0.00065$}
&&\emph{$0.1954  \pm 0.0125$}\\
\emph{$m_Z$}
&&\emph{$0.00233 \pm 0.00045$}
&&\emph{$0.00469 \pm 0.00066$}
&&\emph{$0.0934  \pm 0.0130$}\\
\emph{$m_t$}
&&\emph{$0.00223 \pm 0.00043$}
&&\emph{$0.00449 \pm 0.00064$}
&&\emph{$0.0894  \pm 0.0125$}\\
\emph{$10^9$ $GeV$}
&&\emph{$0.00128 \pm 0.00025$}
&&\emph{$0.00260 \pm 0.00037$}
&&\emph{$0.0519  \pm 0.072$}\\
\emph{$M_X$}
&&\emph{$0.00094 \pm 0.00018$}
&&\emph{$0.00194 \pm 0.00028$}
&&\emph{$0.0387  \pm 0.054$}\\
\hline
&&\emph{c}&&\emph{b}&&\emph{t}\\
\hline
\emph{1 GeV}
&&\emph{$1.506   \pm 0.048$}
&&\emph{$7.18    \pm 0.59$}
&&\emph{$475     \pm 86$} \\
\emph{$m_Z$}
&&\emph{$0.677   \pm 0.061$}
&&\emph{$3.00    \pm 0.11$}
&&\emph{$181     \pm 13$} \\
\emph{$m_t$}
&&\emph{$0.646   \pm 0.059$}
&&\emph{$2.85    \pm 0.11$}
&&\emph{$171     \pm 12$} \\
\emph{$10^9$ $GeV$}
&&\emph{$0.371   \pm 0.033$}
&&\emph{$1.51    \pm 0.06$}
&&\emph{$109     \pm 16$} \\
\emph{$M_X$}
&&\emph{$0.272   \pm 0.024$}
&&\emph{$1.07    \pm 0.04$}
&&\emph{$84      \pm 18$} \\
\hline\hline
\end{tabular}
\end{table}

We use the following values for the mixing angles at low energy scales
(1 $GeV$ to $m_t$)~\cite{pdg00}

\begin{equation}\begin{array}{l}
\sin\theta_{12}=0.2225\pm0.0021\\
\sin\theta_{23}=0.04\pm0.0018\\
\sin\theta_{13}=0.0035\pm0.0009
\end{array}\end{equation}

\noindent
and for the CP-phase we take~\cite{ali99},

\begin{equation}
\delta=(66.5 \pm 30.5)^{\circ}.
\end{equation}

To calculate the mixing angles at $10^9$ $GeV$ and $M_X$, where
$M_X$ is the unification scale of SUSY ($M_X=2\times10^{16}$ $GeV$),
we use the formalism in ref.~\cite{kielanowski00}.
In this work Kielanowski {\it et al.}, find the energy dependence
of the $|V_{ij}|$. 
Recalling that from eq. (11), the mixing angles are related to
the magnitudes of the $V_{ij}$ matrix elements. In particular
we are interested in the magnitudes of $|V_{us}|$, $|V_{cb}|$
and $|V_{ub}|$.
The evolution of these magnitudes are found to be given by,

\begin{equation}
|V_{ub}|^2=\frac{|V_{ub}^0|^2}{|V_{tb}^0|^2[h(t)^2-1]+1}
\end{equation}
\begin{equation}
|V_{cb}|^2=\frac{|V_{cb}^0|^2}{|V_{tb}^0|^2[h(t)^2-1]+1}
\end{equation}

\noindent
where $|V_{ij}^0|$ are the initial values of the CKM matrix
elements and $h(t)$ is expressed as

\[
h(t)=\left(
\frac{1}{1-\frac{3(b+2)}{(4\pi)^2}m_t^2(t_0)\int_{t_o}^t r_g(\tau)d\tau}
\right)^{\frac{c}{2(b+2)}}
\]

\noindent
with
\[
r_g(t)=
\exp{\left(
\frac{2}{(4\pi)^2} \int_{t_0}^t \alpha^u_1(\tau)d\tau
\right)}.
\]

\noindent
In the SM we have,
\[
\alpha^u_1(t)=-\left(\frac{17}{20}g^2_1+\frac{9}{4}g^2_2+8g^2_3\right).
\]

\noindent
with
\[
g_i(t)=\frac{g_i(t_0)}{\sqrt{1-\frac{2b_ig_i^2(t_0)(t-t_0)}{(4\pi)^2}}},
\]

\noindent
and $t=\ln(E/m_t)$. $|V_{us}|$ is obtained from unitarity.
We evaluate at $t_1=\ln(\frac{10^9}{171}) \sim 15.5816$ and 
$t_2=\ln(\frac{2\times10^{16}}{171}) \sim 32.3928$.
We assume there is no running for the $\delta$ phase~\cite{balzereit99}. 
We can collect all the numerical results in Table 2.


\begin{table}[htb]
\caption{Elements of mass matrix $d$}
\label{table:2}
\begin{tabular}{cccc}
\hline\hline
Scale
&\emph{$m_{11}$}
&\emph{$m_{12}e^{i\delta_{12}}$}                      
&\emph{$m_{13}e^{i\delta_{13}}$}\\
\hline
1 $GeV$
&$.0194\pm.0030$
&$(.0412\pm.0023)e^{-i(1.26\pm.5)^{\circ}}$
&$(.0235\pm.0066)e^{-i(70.05\pm31.22)^{\circ}}$\\
$m_Z$
&$.0094\pm.0020$
&$(.0196\pm.0019)e^{-i(1.11\pm.43)^{\circ}}$
&$(.0097\pm.0027)e^{-i(70.58\pm31.31)^{\circ}}$\\
$m_t$
&$.0090\pm.0019$
&$(.0188\pm.0018)e^{-i(1.10\pm.42)^{\circ}}$
&$(.0092\pm.0026)e^{-i(70.61\pm31.32)^{\circ}}$\\
$10^9$ $GeV$
&$.0052\pm.0010$
&$(.0110\pm.0009)e^{-i(1.31\pm.49)^{\circ}}$
&$(.0055\pm.0015)e^{-i(71.04\pm31.38)^{\circ}}$\\
$M_X$
&$.0039\pm.0008$
&$(.0082\pm.0007)e^{-i(15.89\pm6.7)^{\circ}}$
&$(.0048\pm.0013)e^{-i(66.95\pm30.6)^{\circ}}$\\
\hline
&\emph{$m_{22}$}
&\emph{$m_{23}e^{i\delta_{23}}$}
&\emph{$m_{33}$}\\
\hline
1 $GeV$
&$.1987\pm.0309$
&$(.2798\pm.0193)e^{i(.03\pm.01)^{\circ}}$
&$(7.167\pm.6089)$\\
$m_Z$
&$.0936\pm.0161$
&$(.1163\pm.0061)e^{i(.03\pm.01)^{\circ}}$
&$(2.995\pm.1155)$\\
$m_t$
&$.0896\pm.0155$
&$(.1105\pm.0060)e^{i(.03\pm.01)^{\circ}}$
&$(2.845\pm.1155)$\\
$10^9$ $GeV$
&$.0525\pm.0081$
&$(.0671\pm.0035)e^{i(.03\pm.01)^{\circ}}$
&$(1.507\pm.0529)$\\
$M_X$ $GeV$
&$.0400\pm.0062$
&$(.0571\pm.0030)e^{i(.04\pm.01)^{\circ}}$
&$(1.067\pm.0424)$\\
\hline\hline
\end{tabular}
\end{table}

If we consider $m_d$ to be diagonal, instead of $m_u$, a similar
analysis determines the structure of the quark mass matrix $m_u$.
The result of this analysis is shown in Table 3.

\begin{table}[h!]
\caption{Elements of mass matrix $u$}
\label{table:3}
\begin{tabular}{cccc}
\hline\hline
Scale
&\emph{$m_{11}$}
&\emph{$m_{12}e^{i\delta_{12}}$}                      
&\emph{$m_{13}e^{i\delta_{13}}$}\\
\hline
1 $GeV$
&$.0887\pm.0248$
&$(.3923\pm.0216)e^{-i(9.60\pm4.12)^{\circ}}$
&$(1.662\pm.4580)e^{-i(66.92\pm30.60)^{\circ}}$\\
$m_Z$
&$.0387\pm.0074$
&$(.1716\pm.0119)e^{-i(8.25\pm3.32)^{\circ}}$
&$(.6284\pm.1657)e^{-i(67.99\pm30.61)^{\circ}}$\\
$m_t$
&$.0369\pm.0070$
&$(.1635\pm.0114)e^{-i(8.18\pm3.29)^{\circ}}$
&$(.5936\pm.1564)e^{-i(67.0\pm30.61)^{\circ}}$\\
$10^9$ $GeV$
&$.0220\pm.0053$
&$(.1001\pm.0075)e^{-i(11.53\pm4.85)^{\circ}}$
&$(.4340\pm.1138)e^{-i(67.0\pm30.60)^{\circ}}$\\
$M_X$
&$.0171\pm.0050$
&$(.0816\pm.0071)e^{-i(54.1\pm26.6)^{\circ}}$
&$(.4107\pm.1078)e^{-i(66.54\pm30.51)^{\circ}}$\\
\hline
&\emph{$m_{22}$}
&\emph{$m_{23}e^{i\delta_{23}}$}
&\emph{$m_{33}$}\\
\hline
1 $GeV$
&$2.266\pm.563$
&$(18.96\pm2.04)e^{i(.0032\pm.0013)^{\circ}}$
&$474\pm86$\\
$m_Z$
&$.937\pm.118$
&$(7.22\pm.47)e^{i(.0038\pm.0014)^{\circ}}$
&$181\pm13$\\
$m_t$
&$.891\pm.118$
&$(6.82\pm.44)e^{i(.0039\pm.0014)^{\circ}}$
&$171\pm12$\\
$10^9$ $GeV$
&$.591\pm.104$
&$(5.0\pm.43)e^{i(.0035\pm.0013)^{\circ}}$
&$109\pm13$\\
$M_X$
&$.529\pm.116$
&$(4.64\pm.48)e^{i(.0336\pm.0138)^{\circ}}$
&$84\pm13$\\
\hline\hline
\end{tabular}\\
\end{table}

From these numerical expressions for the mass matrices, one
can see that {\it no} element is consistent with zero at
$3\sigma$, for any given scale.
We notice the following hierarchy among the magnitudes of
the mass matrix elements,

\begin{equation}
m_{11} \approx m_{13} < m_{12}< m_{22} < m_{23} \ll m_{33}.
\end{equation}

This hierarchy is similar to the one obtained by 
Fritzsch~\cite{fritzsch97}: $m_{11},m_{12},m_{13}\ll m_{23},m_{22}\ll m_{33}$.
We also have a hierarchy among the phases

\begin{equation}
\delta_{23}\ll\ \delta_{12}\ll \delta_{13}
\end{equation}

\noindent
at low energy scales.
At SUSY scales this hierarchy is no longer valid and we have instead
$\delta_{23} \ll \delta_{12} \approx \delta_{13}$.


\section{Conclusions}

To summarize, 
starting with the weak basis, for which one of the quark mass matrices is
diagonal, we find exact relations that are analysed numerically. We obtain
numerical expressions at different scales, using the mixing angles and quark
masses as input data.
From this analysis we conclude that {\it no} texture is from $1$ $GeV$ up
to $2\times10^{16}$ $GeV$ at $3\sigma$.
We also find the explicit dependence of $\delta_{ij}$ in terms of the
quark masses and the CKM  mixing angles.
Numerical evaluation shows
that we have $\delta_{23}\ll\ \delta_{12}\ll \delta_{13}$ at low scales
and $\delta_{23}\ll\ \delta_{12}\sim \delta_{13}$ at SUSY scales.
We expect in the near future better measurements for the known observables
in the Yukawa sector of the SM. When the measurements of these observables
becomes more precise, we will be able to draw more definitive
conclusions about the existence of a particular texture in the quark mass
matrices.

We would like to thank useful conversations with A. Bouzas, B.
Desai, E. Ma, G. Sanchez-Col\'on and J. Wudka. One of us (RH) wants
to thank the hospitality of the Department of Physics at UCR, where
part of this work was done.
This work was partially supported by Conacyt (M\'exico).


\end{document}